# A Prototype Combination TPC Cherenkov Detector with GEM Readout for Tracking and Particle Identification and its Potential Use at an Electron Ion Collider

Craig Woody[1,a], Babak Azmoun[1], Richard Majka[2], Michael Phipps[1,3], Martin Purschke[1], Nikolai Smirnov[2]

[1]Physics Department, Brookhaven National Laboratory, Upton, NY 11973 USA
[2]Physics Department, Yale University, New Haven, CT 06520 USA
[3]Physics Department, University of Illinois at Urbana-Champaign, Urbana, IL 61801 USA

**Abstract.** A prototype detector is being developed which combines the functions of a Time Projection Chamber for charged particle tracking and a Cherenkov detector for particle identification. The TPC consists of a 10x10x10 cm$^3$ drift volume where the charge is drifted to a 10x10 cm$^2$ triple GEM detector. The charge is measured on a readout plane consisting of 2x10 mm$^2$ chevron pads which provide a spatial resolution ~ 100 μm per point in the chevron direction along with dE/dx information. The Cherenkov portion of the detector consists of a second 10x10 cm$^2$ triple GEM with a photosensitive CsI photocathode on the top layer. This detector measures Cherenkov light produced in the drift gas of the TPC by high velocity particles which are above threshold. $CF_4$ or $CF_4$ mixtures will be used as the drift gas which are highly transparent to UV light and can provide excellent efficiency for detecting Cherenkov photons. The drift gas is also used as the operating gas for both GEM detectors. The prototype detector has been constructed and is currently being tested in the lab with sources and cosmic rays, and additional tests are planned in the future to study the detector in a test beam.

## 1 Introduction

In many nuclear and particle physics experiments, one wants to both track and identify particles in various types of collisions. This is usually done using separate detectors for tracking and particle id. However, this can result in increasing the amount of material inside the sensitive volume of a large detector, and can also use up valuable space inside a magnetic spectrometer. It would therefore be desirable if these two features could be combined together in a single detector without compromising its performance. The detector described here combines the tracking features of a Time Projection Chamber (TPC) and a threshold Cherenkov detector for particle id.

This work is part of a Detector R&D Program for a future Electron Ion Collider that is being planned to be built at either Brookhaven National Lab (eRHIC) or Thomas Jefferson National Lab (MEIC) [1,2]. An EIC would collide beams of electrons with protons and heavy ions at high energies in order to study nucleon structure and QCD over a broad range of x and $Q^2$. A large multipurpose spectrometer would be used to measure deep inelastic electron scattering over a wide range of rapidity and solid angle, and the tracking system for the central detector would consist of a TPC and a precision vertex detector. The combined TPC-Cherenkov detector described here could be used to provide both tracking and particle id information for measuring the scattered electron and separating it from hadrons produced in the central region. An example of such a multipurpose spectrometer for EIC is described in the last section.

## 2 Prototype TPC Cherenkov Detector

We have constructed a prototype detector that combines the tracking features of a Time Projection Chamber (TPC) and a threshold Cherenkov detector for particle id. It utilizes two triple GEM detectors, one for measuring charged tracks in the TPC and another for detecting Cherenkov light produced by high momentum particles in the same gas volume. The charge from tracks passing through the detector is drifted to a triple GEM detector located at the bottom of the drift volume, and Cherenkov light is detected by a second triple GEM which has a photosensitive CsI photocathode deposited on its top layer. This technique for detecting Cherenkov light using a photosensitive GEM detector was also used in the PHENIX Hadron Blind Detector [3] for measuring low mass electron pairs in heavy ion collisions.

Figure 1 shows a 3D model of the detector. It consists of a three sided field cage made of a kapton foil with copper strips and a fourth side made of wires to allow light to pass through to the photosensitive GEM. There is no window between the drift volume and the photocathode which allows for maximum optical transparency. The photosensitive GEM is mounted on a movable stage to allow the distance between it and the TPC to be varied in order to study potential high voltage problems when the two detectors are brought into close proximity to each other. The drift volume is 10x10x10 cm$^3$ and the GEM detectors are 10x10 cm$^2$. The entire assembly is mounted inside a common enclosure and will be filled with a gas mixture that serves as the TPC gas and the operating gas for both GEMs, and also provides a highly UV transparent radiator for Cherenkov light.

[a] Corresponding author: woody@bnl.gov

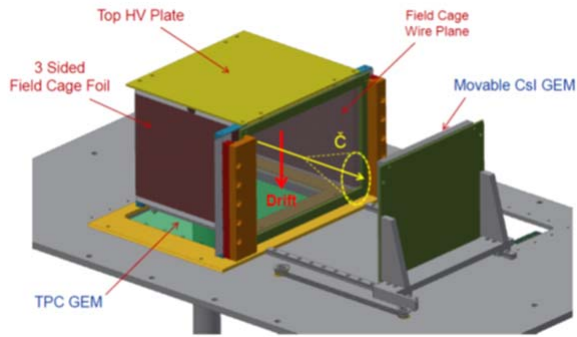

**Figure 1.** 3D model of the prototype TPC/Cherenkov detector.

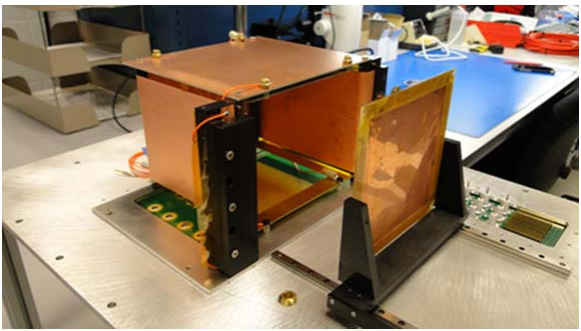

**Figure 2.** Internal components of the actual TPC-Cherenkov prototype detector. The foil on the right is mounted on a movable track such that the distance between the photosensitive GEM and the TPC can be varied.

Figure 2 shows some of the components of the actual detector. The kapton foil field cage consists of 3.9 mm wide copper strips with 0.1 mm gaps in between. There are copper strips on both the front and back of the foil which are displaced by half a strip spacing (2 mm) to improve the field uniformity. For testing the TPC portion of the detector, a fourth side for the field cage made of a similar kapton foil is used as shown in Figure 3. The field cage has been tested up to 1 kV/cm, which is the maximum drift field we expect to use. The wire plane forming the fourth side consists of 75 μm wires spaced 1 mm apart that are connected in groups of four and held at the same potential in order to achieve the same field gradient as the copper strips on the kapton foil. Figure 4 shows the wire plane that is used for the fourth side of the field cage. The separate kapton foil and wire plane can be easily interchanged in order to study the detector as a conventional TPC or in combination with the Cherenkov detector.

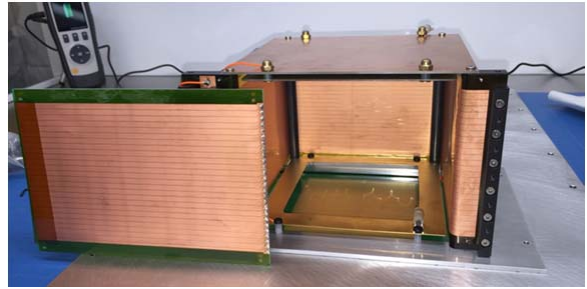

**Figure 3.** Three sided kapton field cage with a separate fourth kapton foil positioned to the side.

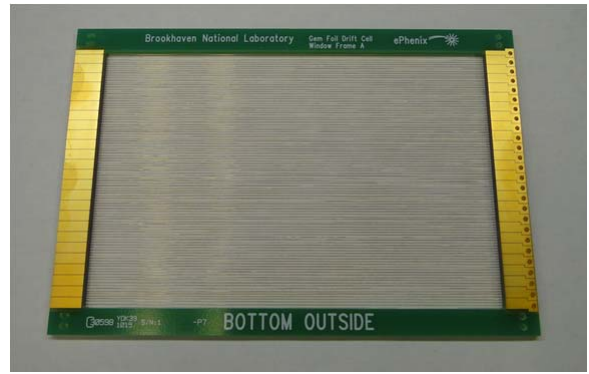

**Figure 4.** Wire plane used as the fourth side of the field cage when operating with the Cherenkov detector.

## 3 Electrostatic Field Simulations

The requirement of an optically transparent side of the field cage and the presence of the photosensitive GEM detector near the drift volume causes some distortion in the drift field of the TPC. This problem was studied using an electrostatic simulation program (ANSYS) in order to determine the magnitude of these distortions. Figure 5 shows the deviation of the nominal electric field vector in the drift volume as a function of distance along the drift direction and the distance perpendicular to the wire plane for the first mesh of the photosensitive GEM at a distance of x = -15 mm. The distortions caused by the wire plane and the presence of the photosensitive GEM are generally less than 1%.

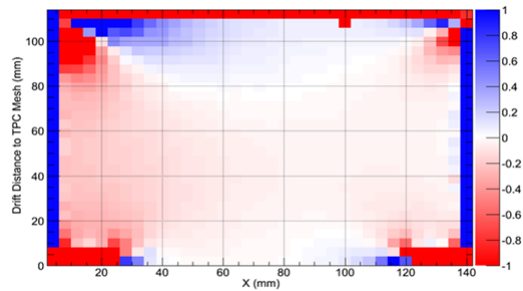

**Figure 5.** Electrostatic simulation showing the deviation in percent of the nominal electric field vector in the drift region as a function of the drift distance and the distance perpendicular to the wire plane of the field cage for the photosensitive GEM located at a distance of x = - 15 mm.

## 4 Initial Tests of the TPC

The TPC portion of the detector has been tested using radioactive sources and cosmic rays. The GEM detector for the TPC and its readout plane are shown in Fig. 6. The readout plane consists of an array of 512 chevron pads, where each pad is 2 mm wide in the chevron direction with a 0.5 mm chevron pitch and is 10 mm long.

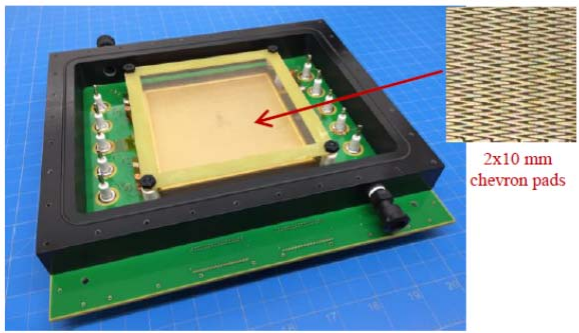

**Figure 6.** GEM detector with chevron readout board for the TPC portion of the prototype.

The chevron pads exhibit an inherent differential non-linearity due to the interleaving zig-zag pattern. However, this non-linearity can be corrected for by measuring the response of the pad in very fine steps. In order to measure this, the detector was configured as a "minidrift" detector with a 1.6 mm drift gap above the GEM [4,5] and scanned across the chevron direction using a highly collimated X-ray source. Figure 7 shows the measured response of the detector as a function of the source position and the differential non-linearity. Figure 8 shows the position resolution before and after the DNL correction. After correcting for the differential non-linearity, the position resolution decreased from 132 μm to 98 μm.

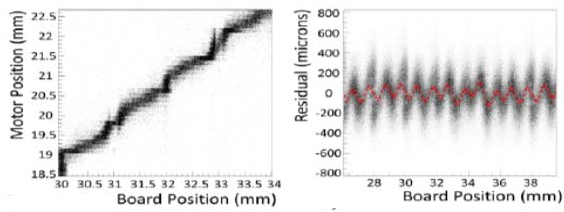

**Figure 7.** Left: Detector response as a function of source position; Right: Differential non-linearity

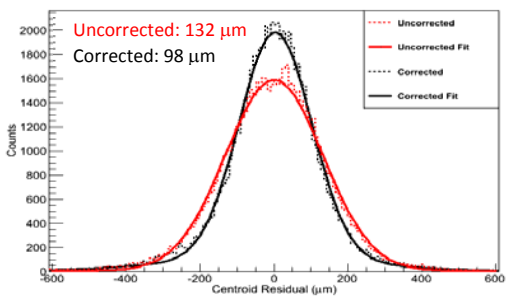

**Figure 8.** Position resolution across the chevron direction before and after applying the differential non-linearity correction.

The detector was also tested with the full field cage using cosmic rays. The readout used the CERN SRS system, which unfortunately is not designed for TPC applications. The system utilizes a 40 MHz clock which samples the preamp pulses every 25 ns, but has a buffer of only 32 samples. This limits the ability to measure the drift time over the entire drift region, and also does not provide an easy way to synchronize the readout with the cosmic ray trigger.

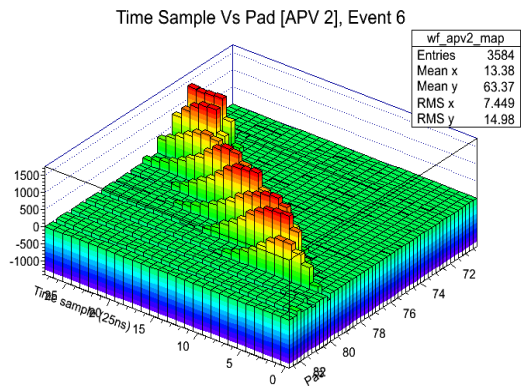

**Figure 9.** GEM detector with chevron readout board for the TPC portion of the prototype.

Nevertheless, we were able to measure tracks in the detector over a range of a few cm in order to demonstrate that the TPC portion of the detector is working. Figure 9 shows the time sampled pulses from a number of pads hit from a cosmic ray traversing the detector at a moderate angle. The data was used to reconstruct the track using a "Time Slice Method", which computes a centroid for all pads hit above a given pulse height threshold in each time bin. This provides a series of points in transverse plane and drift direction that are then used to reconstruct the track. Figure 10 shows an example of a track found in the detector operating with an $Ar/CO_2$ (70/30) gas mixture at a drift field of 0.8 kV/cm, which has a drift velocity ~ 2.2 cm/μsec and corresponds to a distance ~ 14 mm along the drift direction.

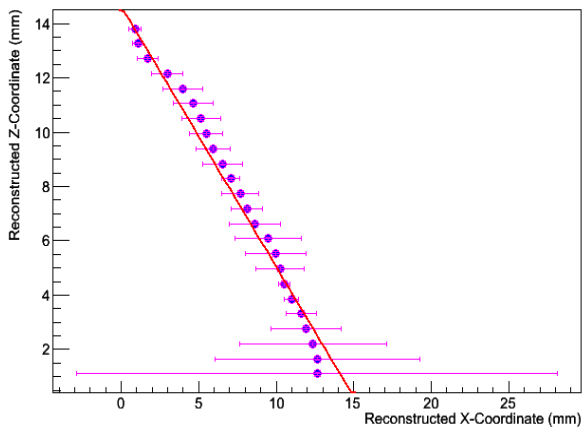

**Figure 10.** Cosmic ray track found in the TPC portion of the detector operating with $Ar/CO_2$ (70/30). The vertical path length corresponds to ~ 14 mm in this gas.

We have also studied the TPC using different gases such as Ar/$CO_2$/$CF_4$ (80/10/10) and Ar/$CH_4$ (80/20, P20). These gases give a somewhat higher drift velocity than Ar/$CO_2$, which is useful for testing the TPC, but are not suitable for operating with the Cherenkov detector. Figure 11 gives a compilation of drift velocities versus drift field for a number of gases we are considering using for the TPC alone and in the combined TPC/Cherenkov detector.

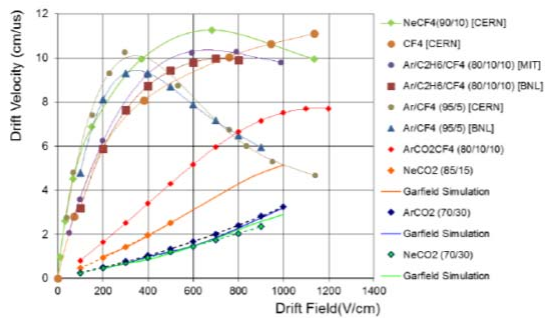

**Figure 11.** Compilation of drift velocities versus drift field for various gases that could be used for operating only the TPC or the combined TPC/Cherenkov detector.

For combined TPC/Cherenkov operation, a gas with high optical transparency into the deep VUV is required in order to have good photon detection efficiency for the photosensitive CsI coated GEM. The PHENIX HBD was operated with pure $CF_4$ and produced a yield of ~ 20 photoelectrons with a radiator length of 50 cm [3]. $CF_4$ is certainly a possibility for the TPC/Cherenkov detector, although a fairly high drift field is required to achieve a high drift velocity, as shown in Fig. 11. Other possibilities are mixtures of $CF_4$ and Ar, both of which have high transparency into the deep VUV as shown in Fig. 12, and can achieve a high drift velocity at a relatively low field as shown in Fig. 11.

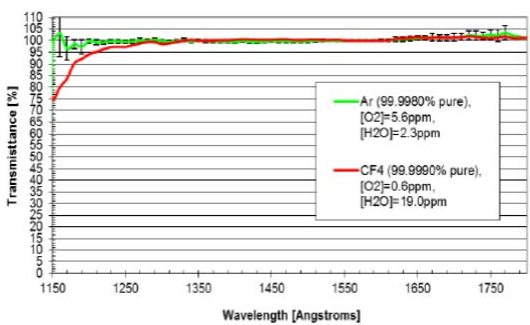

**Figure 12.** Transparency of Ar and $CF_4$ in the deep UV.

## 5 Application at an Electron Ion Collider

A combined TPC/Cherenkov detector would use ionization information for tracking and dE/dx and UV light from particles above the Cherenkov threshold in the same gas volume to provide both tracking and particle id. This would enhance the capabilities of a central TPC tracker for an Electron Ion Collider detector by helping to identify the scattered electron in ep and eA collisions.

Figure 13 shows an example of an EIC detector at eRHIC based on the upgrade of the proposed sPHENIX experiment [6,7]. It consists of a central spectrometer with a TPC tracker surrounded by calorimetry, a high resolution electromagnetic calorimeter in the electron going direction, and a hadronic spectrometer in the hadron going direction. Another example of an eRHIC detector with a central TPC is described in [2]. For the energy range at eRHIC (electrons up to 21.2 GeV/c, protons up to 250 GeV/c and ions up to 100 GeV/A), both electrons and hadrons are scattered into the central region. Therefore, good e/h separation is required over a large range of rapidity (~ |η| < 3), and a TPC tracker with enhanced electron id would be able to provide additional hadron rejection capability in the central region in a single detector.

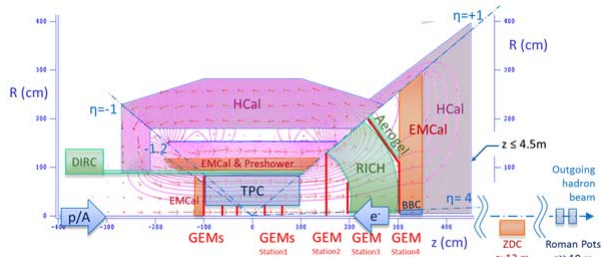

**Figure 13.** An EIC detector based on the sPHENIX experiment at RHIC.

## 5 Summary

A combined TPC/Cherenkov prototype detector is being developed that will provide both tracking and particle id. The detector has been designed and constructed and the TPC portion has been tested with radioactive sources and cosmic rays. The Cherenkov portion will be added next and the combined features of both detectors will be studied in a test beam at Fermilab or SLAC. A TPC/Cherenkov detector such as this would provide enhanced particle identification capabilities for a central tracker at a future Electron Ion Collider.